\documentclass[runningheads,citeauthoryear]{apinv}
\usepackage{epsfig,cite,graphics}
\usepackage[T2A]{fontenc}
\usepackage[cp1251]{inputenc}
\usepackage{lscape}
\usepackage{longtable}

\begin{document}

\title{Long-term optical photometry of the PMS stars V2764 Ori and LkH$\alpha$ 301 in the field of the McNeil's Nebula}
\titlerunning{Long-term photometry of the PMS stars V2764 Ori and LkH$\alpha$ 301}
\author{G. Zidarova\inst{1}, S. Ibryamov\inst{1}, E. Semkov\inst{2}, S. Peneva\inst{2}}
\authorrunning{G. Zidarova et al.}
\tocauthor{G. Zidarova et al.} 
% Command tocautor{} is used by the Latex to give author names 
% to the Contents of the volume (automatically generated)
\institute{Department of Physics and Astronomy, University of Shumen, 115, Universitetska Str., 9700 Shumen, Bulgaria
	\and Institute of Astronomy and National Astronomical Observatory, Bulgarian Academy of Sciences, 72, Tsarigradsko Shose Blvd., 1784 Sofia, Bulgaria
	\newline
	\email{g.zidarova@shu.bg}}
\papertype{Submitted on xxx; Accepted on xxx}	
% Papertype can be "Research report", "Review", "Invited lecture", "Conference talk", 
% "Conference poster", "Lecture at scientific seminar", "Summary of dissertation",  etc.
\maketitle

\begin{abstract}
In this paper, we present results from long-term $V(RI)_{c}$ photometric observations of the pre-main-sequence stars V2764 Ori and LkH$\alpha$ 301, located in the field of the McNeil's Nebula within the Orion star-forming complex.
Our observations were performed in the period from August 2004 to November 2021 with three telescopes and seven different types of CCD cameras.
Photometric observations, especially concerning the long-term behavior of the stars, are missing in the literature.
We present the first photometric monitoring for them, which cover 17 years.
Our data indicate that the variability of both stars is typical for classical T Tauri stars.
\end{abstract}


\section*{1. Introduction}

The McNeil's Nebula is a reflection nebula discovered by McNeil (2004).
It is located in the LDN 1630 molecular cloud within the Orion star-forming complex and associated with the point-like source IRAS 05436-0007 (V1647 Ori).
V1647 Ori is an eruptive young object like FUor or EXor that illuminates the McNeil's Nebula.
The Orion complex is one of the best-known sites of recent star formation containing many molecular clouds, dark and cometary nebulae, Herbig-Haro objects, and pre-main-sequence (PMS) stars.

Photometric variability in PMS stars is one of their main characteristics.
Both classes of PMS stars $-$ the low-mass (M$\leq$2M$_{\odot}$) T Tauri stars (TTSs) and the more massive (2M$_{\odot}$$\leq$M$\leq$8M$_{\odot}$) Herbig Ae/Be stars (HAEBESs) show various types of photometric variability (Herbst et al. 1994).
The TTSs exhibit rapid irregular light variations and emission spectra (Joy 1945).
They are divided into two subgroups: classical T Tauri stars (CTTSs) and weak-line T Tauri stars (WTTSs).
Most of the features that characterize each subgroup suggest that accretion disks surround CTTSs, whereas they must have almost disappeared in WTTSs (see Bertout 1989 and M\'{e}nard \& Bertout 1999).
CTTSs are distinguished from WTTSs by their strong H$\alpha$ emission line and significant infrared and ultraviolet excesses.
The presence of hot and cool spots on the stellar surface, variable accretion activity, circumstellar dust or cloud obscuration events are possible reasons for the variability of TTSs (see Grinin et al. 1991, Herbst et al. 1994 and Ismailov 2005).

The stars included in the present study are located in the field of the McNeil's Nebula.
V2764 Ori was found to be variable by Brice\~{n}o et al. (2004).
In their work, the star was labeled as "V".
LkH$\alpha$ 301 was registered as H$\alpha$-emission star by Herbig \& Kuhi (1963).
Semkov (2004) confirmed that these stars are variables.
Both V2764 Ori and LkH$\alpha$ 301 are X-ray sources (Simon et al. 2004).
Flaherty \& Muzerolle (2008) classified them as CTTSs and determined their stellar parameters $-$ for V2764 Ori $-$ the spectral class K4, the mass 1.56 M$_{\odot}$, the effective temperature 4590 K, the radius 3.04 R$_{\odot}$ and the luminosity 3.576 L$_{\odot}$; for LkH$\alpha$ 301 $-$ the spectral class K2.5, the mass 1.85 M$_{\odot}$, the effective temperature 4815 K, the radius 2.74 R$_{\odot}$ and the luminosity 3.525 L$_{\odot}$.

V2764 Ori and LkH$\alpha$ 301 were studied spectrally (see Flaherty \& Muzerolle 2008, Fang et al. 2009 and Espaillat et al. 2012), but long-term photometry is missing for them.
In this paper, we present the long-term optical light variations of the stars.

\section*{2. Observations and Data reduction}

The $V(RI)_{c}$ photometric observations of the field of the McNeil's Nebula were carried out in the period from August 2004 to November 2021 with the 2-m Ritchey-Chr\'{e}tien-Coud\'{e} (RCC) and the 50/70-cm Schmidt telescopes administered by the Rozhen National Astronomical Observatory (Bulgaria) and the 1.3-m Ritchey-Chr\'{e}tien (RC) telescope administered by the Skinakas Observatory\footnote[1]{Skinakas Observatory is a collaborative project of the University of Crete, the Foundation for Research and Technology, Greece, and the Max-Planck-Institut f{\"u}r Extraterrestrische Physik, Germany.} of the University of Crete (Greece).
Seven different CCD cameras were operated to obtain the observations as follows: VersArray 1300B and Andor iKon-L BEX2-DD on the 2-m RCC telescope, Photometrics CH360 and Andor DZ436-BV on the 1.3-m RC telescope, and SBIG ST-8, SBIG STL-11000M and FLI PL16803 on the 50/70-cm Schmidt telescope.

All frames were taken through a standard Johnson-Cousins ($V(RI)_{c}$) set of filters.
Twilight flat-fields in each filter were obtained each clear evening or morning.
The frames acquired with the cameras on the 2-m RCC and the 1.3-m RC telescopes are bias-frame subtracted and flat-field corrected.
The frames obtained with the cameras on the 50/70-cm Schmidt telescope are dark-frame subtracted and flat-field corrected.

The photometric data were reduced using subroutine \textsc{daophot} in the \textsc{idl} software package.
As a reference sequence we used the $V(RI)_{c}$ comparisons reported in Semkov (2006).
All data were analyzed using the same aperture, which was chosen to have a 5$\arcsec$ radius and background annulus from 10$\arcsec$ to 15$\arcsec$.
The average value of the errors in the reported magnitudes is 0.01-0.02 mag for the $I_{c}$ and $R_{c}$ band data and 0.01-0.03 mag for the $V$ band data.

\section*{3. Results and Discussion}

Figure 1 features an $I_{c}$ band image of the field of the McNeil's Nebula in the LDN 1630, where the positions of the stars from our study and V1647 Ori are marked.

\begin{figure}[]
	\begin{center}
		\centering{\epsfig{file=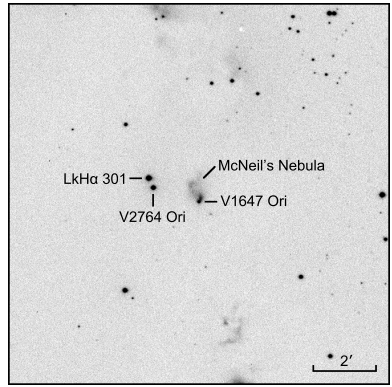, width=5.0cm}}
		\caption[]{An $I_{c}$ band image of the field of the McNeil's Nebula obtained on 05 February 2013 with the 50/70-cm Schmidt telescope. The positions of the stars from our study and the young object V1647 Ori are marked. North is up, and east is left}
		\label{fig1}
	\end{center}
\end{figure}

Results from our long-term CCD observations of V2764 Ori are summarized in Table 1.
The columns in the table contain date ($dd.mm.yyyy$ format) and Julian date (J.D.) of the observations, $I_{c}$$R_{c}$$V$ magnitudes of the star, and telescope and CCD camera used.
The light curves of the star constructed on the basis of our monitoring are displayed in Fig. 2.
The available data suggest that during our observations the brightness of V2764 Ori principally varies around some intermediate level.
The star exhibits irregular variability with different amplitudes in different periods.
In August 2004 and December 2013, short-term decreases in the star's light were registered.

\begin{figure}[]
	\begin{center}
		\centering{\epsfig{file=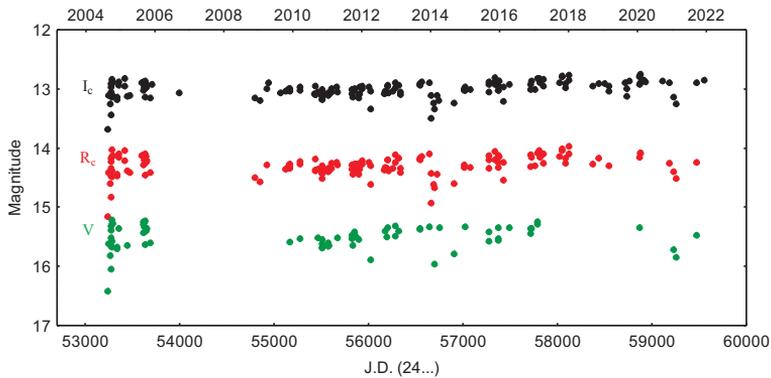, width=10cm}}
		\caption[]{$V(RI)_{c}$ light curves of V2764 Ori for the period August 2004$-$November 2021.}
		\label{fig2}
	\end{center}
\end{figure}

\newpage
{\tiny
	\begin{longtable}{ccccccc|ccccccc}
		\caption{Photometric CCD observations of V2764 Ori.}\\
		\hline\hline
		\noalign{\smallskip}  
		Date & J.D. (24...) & $I_{c}$ & $R_{c}$ & $V$ & Tel & CCD & Date & J.D. (24...) & $I_{c}$ & $R_{c}$ & $V$ & Tel & CCD \\
		\noalign{\smallskip}  
		\hline
		\endfirsthead
		\caption{Continued.}\\
		\hline\hline
		\noalign{\smallskip}  
		Date & J.D. (24...) & $I_{c}$ & $R_{c}$ & $V$ & Tel & CCD & Date & J.D. (24...) & $I_{c}$ & $R_{c}$ & $V$ & Tel & CCD \\
		\noalign{\smallskip}  
		\hline
		\noalign{\smallskip}  
		\endhead
		\hline
		\label{Tab1}
		\endfoot
		\noalign{\smallskip}
18.08.2004	&	53235.723	&	13.67	&	15.15	&	16.41	&	1.3-m	&	Phot	&	03.09.2012	&	56173.572	&	13.03	&	14.31	&	15.40	&	1.3-m	&	AND 	\\
21.08.2004	&	53238.719	&	13.10	&	14.41	&	15.61	&	1.3-m	&	Phot	&	13.09.2012	&	56183.564	&	13.05	&	14.33	&	15.49	&	1.3-m	&	AND 	\\
09.09.2004	&	53257.704	&	13.24	&	14.59	&	15.81	&	1.3-m	&	Phot	&	23.09.2012	&	56193.576	&	12.95	&	14.19	&	15.34	&	1.3-m	&	AND 	\\
18.09.2004	&	53266.733	&	12.96	&	14.22	&	15.37	&	1.3-m	&	Phot	&	23.09.2012	&	56193.597	&	12.93	&	14.19	&	-	&	Sch 	&	FLI	\\
19.09.2004	&	53267.734	&	12.90	&	14.16	&	15.30	&	1.3-m	&	Phot	&	09.10.2012	&	56209.563	&	13.06	&	14.38	&	-	&	Sch 	&	FLI	\\
20.09.2004	&	53268.724	&	13.04	&	14.33	&	15.51	&	1.3-m	&	Phot	&	18.11.2012	&	56249.501	&	13.03	&	14.34	&	-	&	Sch 	&	FLI	\\
21.09.2004	&	53269.672	&	13.12	&	14.46	&	15.65	&	1.3-m	&	Phot	&	13.12.2012	&	56275.457	&	12.88	&	14.11	&	15.31	&	2-m	&	VA	\\
23.09.2004	&	53271.716	&	13.44	&	14.81	&	16.04	&	1.3-m	&	Phot	&	14.12.2012	&	56276.450	&	12.98	&	14.24	&	15.48	&	2-m	&	VA	\\
29.09.2004	&	53277.624	&	13.14	&	14.47	&	15.66	&	1.3-m	&	Phot	&	19.01.2013	&	56312.331	&	12.93	&	14.16	&	15.39	&	2-m	&	VA	\\
30.09.2004	&	53278.619	&	13.09	&	14.38	&	15.57	&	1.3-m	&	Phot	&	04.02.2013	&	56328.303	&	13.09	&	14.40	&	-	&	Sch 	&	FLI	\\
01.10.2004	&	53279.627	&	12.90	&	14.13	&	15.28	&	1.3-m	&	Phot	&	05.02.2013	&	56329.311	&	13.05	&	14.33	&	-	&	Sch 	&	FLI	\\
02.10.2004	&	53280.736	&	12.82	&	14.02	&	15.21	&	1.3-m	&	Phot	&	05.09.2013	&	56540.584	&	12.91	&	14.18	&	15.35	&	Sch 	&	FLI	\\
03.10.2004	&	53281.729	&	12.88	&	14.12	&	15.27	&	1.3-m	&	Phot	&	07.09.2013	&	56542.564	&	12.91	&	14.19	&	-	&	Sch 	&	FLI	\\
19.11.2004	&	53328.498	&	13.17	&	14.46	&	15.69	&	Sch 	&	ST-8	&	08.09.2013	&	56543.593	&	12.93	&	14.16	&	15.37	&	2-m	&	VA	\\
21.11.2004	&	53330.546	&	13.13	&	14.43	&	15.67	&	Sch 	&	ST-8	&	18.09.2013	&	56553.544	&	12.88	&	14.13	&	-	&	1.3-m	&	AND 	\\
08.12.2004	&	53348.359	&	12.93	&	14.14	&	15.35	&	Sch 	&	ST-8	&	09.12.2013	&	56636.381	&	12.89	&	14.09	&	15.33	&	2-m	&	VA	\\
10.12.2004	&	53350.390	&	12.87	&	14.10	&	-	&	Sch 	&	ST-8	&	28.12.2013	&	56655.413	&	13.48	&	14.92	&	-	&	Sch 	&	FLI	\\
10.02.2005	&	53412.366	&	12.81	&	14.03	&	-	&	Sch 	&	ST-8	&	29.12.2013	&	56656.404	&	13.10	&	14.41	&	-	&	Sch 	&	FLI	\\
11.02.2005	&	53413.350	&	12.95	&	14.21	&	-	&	Sch 	&	ST-8	&	23.01.2014	&	56681.349	&	13.23	&	14.61	&	-	&	Sch 	&	FLI	\\
12.03.2005	&	53442.273	&	13.11	&	14.38	&	15.63	&	2-m	&	VA	&	05.02.2014	&	56694.316	&	13.33	&	14.67	&	15.96	&	2-m	&	VA	\\
03.04.2005	&	53464.242	&	13.11	&	14.40	&	-	&	Sch 	&	ST-8	&	26.02.2014	&	56715.250	&	13.10	&	14.43	&	-	&	Sch 	&	FLI	\\
14.08.2005	&	53596.730	&	12.89	&	14.11	&	-	&	1.3-m	&	Phot	&	22.03.2014	&	56739.222	&	13.18	&	-	&	-	&	Sch 	&	FLI	\\
27.08.2005	&	53609.713	&	13.02	&	14.28	&	15.43	&	1.3-m	&	Phot	&	30.03.2014	&	56747.311	&	-	&	-	&	15.34	&	2-m	&	VA	\\
28.08.2005	&	53610.711	&	12.88	&	14.10	&	15.23	&	1.3-m	&	Phot	&	30.08.2014	&	56899.567	&	13.23	&	14.59	&	15.79	&	1.3-m	&	AND 	\\
29.08.2005	&	53611.708	&	12.88	&	14.13	&	15.28	&	1.3-m	&	Phot	&	13.12.2014	&	57005.488	&	13.02	&	14.32	&	-	&	Sch 	&	FLI	\\
03.09.2005	&	53616.724	&	12.96	&	14.20	&	15.32	&	1.3-m	&	Phot	&	15.12.2014	&	57006.601	&	12.99	&	14.28	&	-	&	Sch 	&	FLI	\\
10.09.2005	&	53623.724	&	12.91	&	-	&	15.26	&	1.3-m	&	Phot	&	24.12.2014	&	57016.438	&	12.95	&	-	&	15.33	&	2-m	&	VA	\\
11.09.2005	&	53624.724	&	12.86	&	14.09	&	15.23	&	1.3-m	&	Phot	&	25.12.2014	&	57017.371	&	13.02	&	-	&	-	&	2-m	&	VA	\\
15.09.2005	&	53628.702	&	13.12	&	14.45	&	15.62	&	1.3-m	&	Phot	&	20.02.2015	&	57074.281	&	13.00	&	14.32	&	-	&	Sch 	&	FLI	\\
19.09.2005	&	53632.688	&	12.94	&	14.20	&	-	&	1.3-m	&	Phot	&	02.09.2015	&	57267.570	&	13.04	&	14.34	&	15.57	&	1.3-m	&	AND 	\\
20.09.2005	&	53633.706	&	12.97	&	14.23	&	15.39	&	1.3-m	&	Phot	&	03.09.2015	&	57268.593	&	12.92	&	14.19	&	15.40	&	1.3-m	&	AND 	\\
25.09.2005	&	53638.699	&	12.91	&	14.13	&	-	&	1.3-m	&	Phot	&	04.09.2015	&	57269.581	&	12.87	&	14.12	&	-	&	1.3-m	&	AND 	\\
03.10.2005	&	53646.700	&	12.94	&	14.20	&	15.35	&	1.3-m	&	Phot	&	03.11.2015	&	57330.281	&	12.91	&	14.20	&	-	&	Sch 	&	FLI	\\
03.11.2005	&	53678.396	&	13.15	&	14.41	&	15.60	&	2-m	&	VA	&	05.11.2015	&	57331.511	&	12.80	&	14.04	&	-	&	Sch 	&	FLI	\\
26.11.2005	&	53701.340	&	12.92	&	-	&	-	&	Sch 	&	ST-8	&	06.11.2015	&	57333.444	&	12.88	&	-	&	-	&	Sch 	&	FLI	\\
10.09.2006	&	53988.592	&	13.06	&	-	&	-	&	1.3-m	&	Phot	&	07.11.2015	&	57333.502	&	12.88	&	-	&	-	&	Sch 	&	FLI	\\
20.11.2008	&	54791.385	&	13.14	&	14.49	&	-	&	Sch 	&	STL-11	&	08.11.2015	&	57334.501	&	12.87	&	-	&	-	&	Sch 	&	FLI	\\
11.01.2009	&	54843.261	&	13.18	&	14.57	&	-	&	Sch 	&	STL-11	&	12.12.2015	&	57369.458	&	12.94	&	14.12	&	15.34	&	2-m	&	VA	\\
24.03.2009	&	54915.260	&	12.99	&	14.28	&	-	&	Sch 	&	STL-11	&	13.12.2015	&	57370.436	&	13.02	&	14.32	&	15.55	&	2-m	&	VA	\\
16.04.2009	&	54938.268	&	12.89	&	-	&	-	&	Sch 	&	STL-11	&	14.12.2015	&	57371.436	&	13.02	&	14.23	&	15.52	&	2-m	&	VA	\\
22.08.2009	&	55065.559	&	13.06	&	-	&	-	&	Sch 	&	FLI	&	15.12.2015	&	57372.435	&	12.86	&	14.11	&	-	&	Sch 	&	FLI	\\
09.10.2009	&	55114.450	&	13.03	&	14.35	&	-	&	Sch 	&	FLI	&	17.12.2015	&	57374.398	&	12.86	&	14.13	&	-	&	Sch 	&	FLI	\\
28.10.2009	&	55133.417	&	13.02	&	14.34	&	-	&	Sch 	&	FLI	&	06.02.2016	&	57425.312	&	12.95	&	14.23	&	-	&	Sch 	&	FLI	\\
19.11.2009	&	55155.408	&	12.99	&	14.28	&	-	&	Sch 	&	FLI	&	07.02.2016	&	57426.302	&	13.20	&	14.54	&	-	&	Sch 	&	FLI	\\
20.11.2009	&	55156.470	&	12.99	&	14.27	&	-	&	Sch 	&	FLI	&	05.04.2016	&	57484.260	&	12.91	&	-	&	15.34	&	2-m	&	VA	\\
21.11.2009	&	55157.471	&	13.01	&	14.31	&	-	&	Sch 	&	FLI	&	22.11.2016	&	57714.500	&	12.92	&	14.10	&	15.35	&	2-m	&	VA	\\
25.11.2009	&	55161.496	&	13.03	&	14.34	&	15.57	&	2-m	&	VA	&	22.11.2016	&	57715.490	&	13.00	&	14.31	&	15.44	&	2-m	&	VA	\\
11.03.2010	&	55267.275	&	13.01	&	14.26	&	15.52	&	2-m	&	VA	&	23.11.2016	&	57716.479	&	12.87	&	14.15	&	15.35	&	2-m	&	VA	\\
12.03.2010	&	55268.321	&	12.96	&	14.22	&	-	&	2-m	&	VA	&	02.01.2017	&	57756.370	&	13.00	&	14.29	&	-	&	Sch 	&	FLI	\\
12.08.2010	&	55420.614	&	13.07	&	-	&	-	&	1.3-m	&	AND 	&	27.01.2017	&	57781.304	&	12.81	&	14.08	&	-	&	Sch 	&	FLI	\\
20.08.2010	&	55428.606	&	12.94	&	14.18	&	-	&	1.3-m	&	AND 	&	28.01.2017	&	57782.376	&	-	&	-	&	15.28	&	2-m	&	VA	\\
24.08.2010	&	55432.586	&	13.09	&	14.38	&	-	&	1.3-m	&	AND 	&	01.02.2017	&	57786.333	&	-	&	-	&	15.24	&	2-m	&	VA	\\
25.08.2010	&	55433.594	&	13.09	&	14.37	&	-	&	1.3-m	&	AND 	&	15.02.2017	&	57800.275	&	12.81	&	14.04	&	-	&	Sch 	&	FLI	\\
20.09.2010	&	55459.636	&	13.05	&	14.35	&	15.50	&	1.3-m	&	AND 	&	16.02.2017	&	57801.305	&	12.83	&	14.07	&	-	&	Sch 	&	FLI	\\
30.10.2010	&	55499.524	&	13.07	&	14.36	&	15.61	&	2-m	&	VA	&	28.02.2017	&	57813.257	&	12.88	&	14.15	&	-	&	Sch 	&	FLI	\\
31.10.2010	&	55500.503	&	13.10	&	14.41	&	15.68	&	2-m	&	VA	&	04.03.2017	&	57817.268	&	12.85	&	14.11	&	-	&	Sch 	&	FLI	\\
01.11.2010	&	55501.501	&	13.15	&	14.42	&	15.66	&	2-m	&	VA	&	01.04.2017	&	57845.249	&	12.83	&	14.09	&	-	&	Sch 	&	FLI	\\
01.11.2010	&	55501.546	&	13.10	&	14.40	&	-	&	Sch 	&	FLI	&	02.04.2017	&	57846.255	&	12.95	&	14.25	&	-	&	Sch 	&	FLI	\\
02.11.2010	&	55502.569	&	13.04	&	14.32	&	15.54	&	2-m	&	VA	&	17.09.2017	&	58013.548	&	12.88	&	14.14	&	-	&	Sch 	&	FLI	\\
04.11.2010	&	55505.472	&	13.00	&	14.28	&	-	&	Sch 	&	FLI	&	17.10.2017	&	58043.523	&	12.79	&	14.03	&	-	&	Sch 	&	FLI	\\
06.11.2010	&	55506.501	&	13.11	&	14.42	&	-	&	Sch 	&	FLI	&	18.10.2017	&	58044.595	&	12.77	&	14.00	&	-	&	Sch 	&	FLI	\\
07.11.2010	&	55507.511	&	13.17	&	14.51	&	-	&	Sch 	&	FLI	&	22.11.2017	&	58080.483	&	12.90	&	14.17	&	-	&	Sch 	&	FLI	\\
01.01.2011	&	55563.420	&	13.06	&	14.36	&	-	&	Sch 	&	FLI	&	23.11.2017	&	58081.486	&	12.98	&	14.25	&	-	&	Sch 	&	FLI	\\
06.01.2011	&	55568.388	&	13.10	&	14.39	&	15.66	&	2-m	&	VA	&	25.12.2017	&	58113.402	&	12.76	&	13.96	&	-	&	Sch 	&	FLI	\\
08.01.2011	&	55570.273	&	13.08	&	14.34	&	15.60	&	2-m	&	VA	&	26.12.2017	&	58114.397	&	12.85	&	14.10	&	-	&	Sch 	&	FLI	\\
09.01.2011	&	55571.378	&	13.09	&	14.37	&	15.63	&	2-m	&	VA	&	03.09.2018	&	58364.578	&	12.95	&	14.26	&	-	&	Sch 	&	FLI	\\
06.02.2011	&	55599.321	&	13.05	&	14.33	&	-	&	Sch 	&	FLI	&	05.11.2018	&	58428.435	&	12.90	&	14.16	&	-	&	Sch 	&	FLI	\\
07.02.2011	&	55600.299	&	12.99	&	14.25	&	-	&	Sch 	&	FLI	&	12.01.2019	&	58496.316	&	12.90	&	-	&	-	&	Sch 	&	FLI	\\
04.04.2011	&	55656.253	&	12.96	&	14.24	&	-	&	Sch 	&	FLI	&	28.02.2019	&	58543.232	&	12.95	&	-	&	-	&	Sch 	&	FLI	\\
09.04.2011	&	55661.260	&	13.04	&	14.28	&	15.51	&	2-m	&	VA	&	01.03.2019	&	58544.250	&	13.03	&	14.29	&	-	&	2-m	&	AND 	\\
11.09.2011	&	55815.564	&	13.01	&	14.27	&	15.46	&	1.3-m	&	AND 	&	12.08.2019	&	58707.587	&	12.89	&	-	&	-	&	Sch 	&	FLI	\\
12.09.2011	&	55816.591	&	13.07	&	14.34	&	15.52	&	1.3-m	&	AND 	&	01.09.2019	&	58727.587	&	12.99	&	-	&	-	&	2-m	&	AND 	\\
20.09.2011	&	55824.537	&	13.13	&	14.43	&	15.64	&	1.3-m	&	AND 	&	03.09.2019	&	58729.589	&	13.12	&	-	&	-	&	2-m	&	AND 	\\
08.10.2011	&	55842.510	&	12.99	&	14.26	&	15.41	&	1.3-m	&	AND 	&	02.10.2019	&	58758.539	&	12.86	&	-	&	-	&	Sch 	&	FLI	\\
14.10.2011	&	55848.501	&	13.02	&	14.28	&	15.42	&	1.3-m	&	AND 	&	15.01.2020	&	58864.358	&	12.78	&	-	&	-	&	Sch 	&	FLI	\\
30.10.2011	&	55865.483	&	-	&	14.37	&	-	&	2-m	&	VA	&	16.01.2020	&	58865.361	&	12.77	&	-	&	-	&	Sch 	&	FLI	\\
01.11.2011	&	55866.521	&	-	&	14.26	&	15.50	&	2-m	&	VA	&	18.01.2020	&	58867.372	&	12.91	&	14.15	&	15.33	&	2-m	&	AND 	\\
26.11.2011	&	55892.474	&	13.09	&	14.33	&	15.53	&	2-m	&	VA	&	20.01.2020	&	58869.361	&	12.89	&	14.07	&	-	&	2-m	&	AND 	\\
27.11.2011	&	55893.417	&	13.14	&	14.43	&	-	&	Sch 	&	FLI	&	21.01.2020	&	58870.366	&	12.75	&	14.07	&	-	&	Sch 	&	FLI	\\
29.11.2011	&	55895.486	&	13.08	&	14.35	&	-	&	Sch 	&	FLI	&	29.02.2020	&	58909.278	&	12.85	&	-	&	-	&	Sch 	&	FLI	\\
30.11.2011	&	55896.443	&	12.96	&	14.24	&	-	&	Sch 	&	FLI	&	21.03.2020	&	58930.318	&	12.87	&	-	&	-	&	Sch 	&	FLI	\\
29.12.2011	&	55925.450	&	12.94	&	14.20	&	-	&	Sch 	&	FLI	&	13.09.2020	&	59105.602	&	12.86	&	-	&	-	&	Sch 	&	FLI	\\
01.01.2012	&	55928.381	&	12.99	&	14.26	&	-	&	Sch 	&	FLI	&	22.11.2020	&	59176.451	&	12.88	&	14.24	&	-	&	Sch 	&	FLI	\\
16.03.2012	&	56003.258	&	12.96	&	14.24	&	-	&	Sch 	&	FLI	&	05.01.2021	&	59220.339	&	13.12	&	14.39	&	15.71	&	2-m	&	AND 	\\
26.03.2012	&	56013.259	&	13.33	&	14.61	&	15.88	&	2-m	&	VA	&	04.02.2021	&	59250.350	&	13.24	&	14.51	&	15.85	&	2-m	&	AND 	\\
28.03.2012	&	56015.310	&	13.03	&	14.29	&	-	&	2-m	&	VA	&	11.09.2021	&	59468.598	&	12.88	&	14.23	&	15.46	&	Sch 	&	FLI	\\
21.08.2012	&	56160.582	&	13.07	&	14.36	&	-	&	1.3-m	&	AND 	&	30.11.2021	&	59549.451	&	12.85	&	-	&	-	&	Sch 	&	FLI	\\
		\hline \hline
\end{longtable}}

The brightness variation of V2764 Ori during the whole period of our monitoring is in the range 12.75-13.67 mag for the $I_{c}$ band, 13.96-15.15 mag for the $R_{c}$ band and 15.21-16.41 mag for the $V$ band.
The registered amplitudes of the light variations of the star during the same period are 0.92 mag for the $I_{c}$ band, 1.19 mag for the $R_{c}$ band and 1.20 mag for the $V$ band.
Variability with such amplitudes is observed in CTTSs and it can be explained by variations in the mass accretion rate and rotational modulation of a spot or group of spots on the stellar surface.

The measured color indices ($V-I_{c}$ and $V-R_{c}$) versus the $V$ magnitude of V2764 Ori are depicted in Fig. 3.
As can be seen, the star becomes redder as it fades.
Such color variation is typical for T Tauri variables, whose variability is produced by the presence of a spot or group of spots on the star's surface, as well as by the small irregular obscuration by the circumstellar material.

\begin{figure}
	\begin{center}
		\includegraphics[width=6cm]{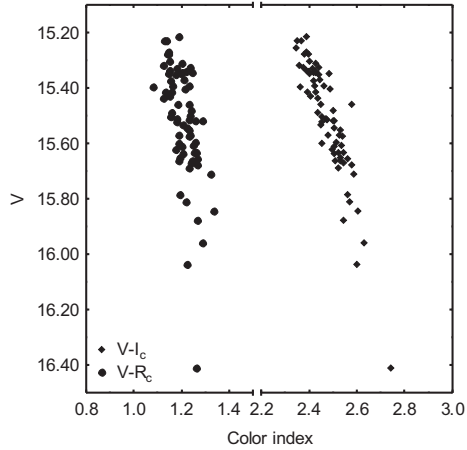}
		\caption{Color indices $V-I_{c}$ and $V-R_{c}$ versus the $V$ magnitude of V2764 Ori.}\label{Fig3}
	\end{center}
\end{figure}

LkH$\alpha$ 301 is located at 20$\arcsec$ from V2764 Ori.
Results from our long-term photometric monitoring of the star are summarized in Table 2.
The columns have the same content as in Table 1.
Figure 4 shows $V(RI)_{c}$ light curves of LkH$\alpha$ 301 from our observations.
As can be seen, like the star V2764 Ori, LkH$\alpha$ 301 exhibits irregular light variations and its brightness varies around some intermediate level.
In September 2004 and September 2019, two short-term decreases in the star's brightness were registered.
During our monitoring the light variation of LkH$\alpha$ 301 is in the range 12.31-13.01 mag for the $I_{c}$ band, 13.38-14.27 mag for the $R_{c}$ band and 14.46-15.43 mag for the $V$ band.
The observed amplitudes are $\Delta I_{c}$=0.70 mag, $\Delta R_{c}$=0.89 mag and $\Delta V$=0.97 mag, which are typical for CTTSs.

\newpage
{\tiny
	\begin{longtable}{ccccccc|ccccccc}
		\caption{Photometric CCD observations of LkH$\alpha$ 301.}\\
		\hline\hline
		\noalign{\smallskip}  
		Date & J.D. (24...) & $I_{c}$ & $R_{c}$ & $V$ & Tel & CCD & Date & J.D. (24...) & $I_{c}$ & $R_{c}$ & $V$ & Tel & CCD \\
		\noalign{\smallskip}  
		\hline
		\endfirsthead
		\caption{Continued.}\\
		\hline\hline
		\noalign{\smallskip}  
		Date & J.D. (24...) & $I_{c}$ & $R_{c}$ & $V$ & Tel & CCD & Date & J.D. (24...) & $I_{c}$ & $R_{c}$ & $V$ & Tel & CCD \\
		\noalign{\smallskip}  
		\hline
		\noalign{\smallskip}  
		\endhead
		\hline
		\label{Tab2}
		\endfoot
		\noalign{\smallskip}
18.08.2004	&	53235.723	&	12.61	&	13.78	&	14.89	&	1.3-m	&	Phot	&	03.09.2012	&	56173.572	&	12.41	&	13.48	&	14.46	&	1.3-m	&	AND 	\\
21.08.2004	&	53238.719	&	12.67	&	13.86	&	15.00	&	1.3-m	&	Phot	&	13.09.2012	&	56183.564	&	12.41	&	13.45	&	14.49	&	1.3-m	&	AND 	\\
09.09.2004	&	53257.704	&	13.01	&	14.27	&	15.43	&	1.3-m	&	Phot	&	23.09.2012	&	56193.576	&	12.45	&	13.50	&	14.54	&	1.3-m	&	AND 	\\
18.09.2004	&	53266.733	&	12.59	&	13.72	&	14.81	&	1.3-m	&	Phot	&	23.09.2012	&	56193.597	&	12.44	&	13.51	&	-	&	Sch 	&	FLI	\\
19.09.2004	&	53267.734	&	12.95	&	14.19	&	15.34	&	1.3-m	&	Phot	&	09.10.2012	&	56209.563	&	12.34	&	13.44	&	-	&	Sch 	&	FLI	\\
20.09.2004	&	53268.724	&	12.62	&	13.78	&	14.87	&	1.3-m	&	Phot	&	18.11.2012	&	56249.501	&	12.38	&	13.48	&	-	&	Sch 	&	FLI	\\
21.09.2004	&	53269.672	&	12.55	&	13.69	&	14.76	&	1.3-m	&	Phot	&	13.12.2012	&	56275.457	&	12.46	&	13.58	&	14.68	&	2-m	&	VA	\\
23.09.2004	&	53271.716	&	12.57	&	13.71	&	14.79	&	1.3-m	&	Phot	&	14.12.2012	&	56276.450	&	12.46	&	-	&	14.72	&	2-m	&	VA	\\
29.09.2004	&	53277.624	&	12.47	&	13.61	&	14.73	&	1.3-m	&	Phot	&	19.01.2013	&	56312.331	&	12.45	&	13.55	&	-	&	2-m	&	VA	\\
30.09.2004	&	53278.619	&	12.45	&	13.55	&	14.64	&	1.3-m	&	Phot	&	04.02.2013	&	56328.303	&	12.56	&	13.71	&	-	&	Sch 	&	FLI	\\
01.10.2004	&	53279.627	&	12.50	&	13.61	&	14.69	&	1.3-m	&	Phot	&	05.02.2013	&	56329.311	&	12.40	&	13.49	&	-	&	Sch 	&	FLI	\\
02.10.2004	&	53280.736	&	12.48	&	13.58	&	14.69	&	1.3-m	&	Phot	&	05.09.2013	&	56540.584	&	12.43	&	13.53	&	14.61	&	Sch 	&	FLI	\\
03.10.2004	&	53281.729	&	12.49	&	13.62	&	14.69	&	1.3-m	&	Phot	&	07.09.2013	&	56542.564	&	12.42	&	13.54	&	-	&	Sch 	&	FLI	\\
19.11.2004	&	53328.498	&	12.56	&	13.69	&	14.86	&	Sch 	&	ST-8	&	08.09.2013	&	56543.593	&	12.45	&	13.55	&	14.63	&	2-m	&	VA	\\
21.11.2004	&	53330.546	&	12.54	&	13.64	&	14.79	&	Sch 	&	ST-8	&	18.09.2013	&	56553.544	&	12.57	&	13.73	&	-	&	1.3-m	&	AND 	\\
08.12.2004	&	53348.359	&	12.45	&	13.57	&	14.73	&	Sch 	&	ST-8	&	09.12.2013	&	56636.381	&	12.53	&	13.63	&	14.75	&	2-m	&	VA	\\
10.12.2004	&	53350.390	&	12.54	&	13.66	&	-	&	Sch 	&	ST-8	&	28.12.2013	&	56655.413	&	12.42	&	13.52	&	-	&	Sch 	&	FLI	\\
10.02.2005	&	53412.366	&	12.41	&	13.51	&	-	&	Sch 	&	ST-8	&	29.12.2013	&	56656.404	&	12.43	&	13.51	&	-	&	Sch 	&	FLI	\\
11.02.2005	&	53413.350	&	12.43	&	13.55	&	-	&	Sch 	&	ST-8	&	23.01.2014	&	56681.349	&	12.46	&	13.58	&	-	&	Sch 	&	FLI	\\
12.03.2005	&	53442.273	&	12.47	&	13.60	&	14.72	&	2-m	&	VA	&	26.02.2014	&	56715.250	&	12.45	&	13.57	&	-	&	Sch 	&	FLI	\\
03.04.2005	&	53464.242	&	12.49	&	13.62	&	-	&	Sch 	&	ST-8	&	22.03.2014	&	56739.222	&	12.40	&	-	&	-	&	Sch 	&	FLI	\\
14.08.2005	&	53596.730	&	12.45	&	13.61	&	-	&	1.3-m	&	Phot	&	30.03.2014	&	56747.311	&	-	&	-	&	14.69	&	2-m	&	VA	\\
27.08.2005	&	53609.713	&	12.45	&	13.59	&	14.67	&	1.3-m	&	Phot	&	30.08.2014	&	56899.567	&	12.45	&	13.57	&	14.61	&	1.3-m	&	AND 	\\
28.08.2005	&	53610.711	&	12.44	&	13.59	&	14.67	&	1.3-m	&	Phot	&	13.12.2014	&	57005.488	&	12.43	&	13.53	&	-	&	Sch 	&	FLI	\\
29.08.2005	&	53611.708	&	12.37	&	13.51	&	14.58	&	1.3-m	&	Phot	&	15.12.2014	&	57006.601	&	12.42	&	13.49	&	-	&	Sch 	&	FLI	\\
03.09.2005	&	53616.724	&	-	&	13.56	&	14.62	&	1.3-m	&	Phot	&	24.12.2014	&	57016.438	&	12.50	&	13.57	&	14.69	&	2-m	&	VA	\\
10.09.2005	&	53623.724	&	12.46	&	-	&	14.67	&	1.3-m	&	Phot	&	25.12.2014	&	57017.371	&	12.43	&	-	&	-	&	2-m	&	VA	\\
11.09.2005	&	53624.724	&	-	&	13.63	&	14.71	&	1.3-m	&	Phot	&	20.02.2015	&	57074.281	&	12.40	&	13.49	&	-	&	Sch 	&	FLI	\\
15.09.2005	&	53628.702	&	12.41	&	13.53	&	14.57	&	1.3-m	&	Phot	&	02.09.2015	&	57267.570	&	12.46	&	13.53	&	14.63	&	1.3-m	&	AND 	\\
19.09.2005	&	53632.688	&	12.38	&	13.51	&	-	&	1.3-m	&	Phot	&	03.09.2015	&	57268.593	&	12.47	&	13.57	&	14.66	&	1.3-m	&	AND 	\\
20.09.2005	&	53633.706	&	12.41	&	13.52	&	14.57	&	1.3-m	&	Phot	&	04.09.2015	&	57269.581	&	12.46	&	13.56	&	-	&	1.3-m	&	AND 	\\
25.09.2005	&	53638.699	&	12.46	&	13.59	&	-	&	1.3-m	&	Phot	&	03.11.2015	&	57330.281	&	12.43	&	13.54	&	-	&	Sch 	&	FLI	\\
03.10.2005	&	53646.700	&	12.45	&	13.61	&	14.68	&	1.3-m	&	Phot	&	05.11.2015	&	57331.511	&	12.40	&	13.50	&	-	&	Sch 	&	FLI	\\
03.11.2005	&	53678.396	&	12.50	&	13.67	&	14.76	&	2-m	&	VA	&	06.11.2015	&	57333.444	&	12.39	&	-	&	-	&	Sch 	&	FLI	\\
26.11.2005	&	53701.340	&	12.41	&	-	&	-	&	Sch 	&	ST-8	&	07.11.2015	&	57333.502	&	12.40	&	-	&	-	&	Sch 	&	FLI	\\
10.09.2006	&	53988.592	&	12.44	&	-	&	-	&	1.3-m	&	Phot	&	08.11.2015	&	57334.501	&	12.49	&	-	&	-	&	Sch 	&	FLI	\\
20.11.2008	&	54791.385	&	12.45	&	13.62	&	-	&	Sch 	&	STL-11	&	12.12.2015	&	57369.458	&	12.38	&	13.44	&	14.53	&	2-m	&	VA	\\
11.01.2009	&	54843.261	&	12.41	&	13.57	&	-	&	Sch 	&	STL-11	&	13.12.2015	&	57370.436	&	-	&	-	&	14.63	&	2-m	&	VA	\\
24.03.2009	&	54915.260	&	12.42	&	13.59	&	-	&	Sch 	&	STL-11	&	14.12.2015	&	57371.436	&	12.41	&	13.48	&	14.61	&	2-m	&	VA	\\
16.04.2009	&	54938.268	&	12.57	&	-	&	-	&	Sch 	&	STL-11	&	15.12.2015	&	57372.435	&	12.43	&	13.55	&	-	&	Sch 	&	FLI	\\
22.08.2009	&	55065.559	&	12.33	&	-	&	-	&	Sch 	&	STL-11	&	17.12.2015	&	57374.398	&	12.39	&	13.49	&	-	&	Sch 	&	FLI	\\
09.10.2009	&	55114.450	&	12.34	&	13.45	&	-	&	Sch 	&	FLI	&	06.02.2016	&	57425.312	&	12.48	&	13.60	&	-	&	Sch 	&	FLI	\\
28.10.2009	&	55133.417	&	12.42	&	13.53	&	-	&	Sch 	&	FLI	&	07.02.2016	&	57426.302	&	12.44	&	13.53	&	-	&	Sch 	&	FLI	\\
19.11.2009	&	55155.408	&	12.36	&	13.45	&	-	&	Sch 	&	FLI	&	05.04.2016	&	57484.260	&	12.48	&	-	&	14.65	&	2-m	&	VA	\\
20.11.2009	&	55156.470	&	12.40	&	13.52	&	-	&	Sch 	&	FLI	&	22.11.2016	&	57714.500	&	-	&	-	&	14.61	&	2-m	&	VA	\\
21.11.2009	&	55157.471	&	12.31	&	13.38	&	-	&	Sch 	&	FLI	&	22.11.2016	&	57715.490	&	12.44	&	-	&	14.61	&	2-m	&	VA	\\
25.11.2009	&	55161.496	&	12.42	&	13.58	&	14.68	&	2-m	&	VA	&	23.11.2016	&	57716.479	&	12.42	&	13.59	&	14.69	&	2-m	&	VA	\\
11.03.2010	&	55267.275	&	12.43	&	13.54	&	14.65	&	2-m	&	VA	&	02.01.2017	&	57756.370	&	12.44	&	13.54	&	-	&	Sch 	&	FLI	\\
12.03.2010	&	55268.321	&	12.40	&	13.54	&	-	&	2-m	&	VA	&	27.01.2017	&	57781.304	&	12.40	&	13.51	&	-	&	Sch 	&	FLI	\\
12.08.2010	&	55420.614	&	12.44	&	-	&	-	&	1.3-m	&	AND 	&	01.02.2017	&	57786.333	&	-	&	-	&	14.62	&	2-m	&	VA	\\
20.08.2010	&	55428.606	&	12.37	&	13.43	&	-	&	1.3-m	&	AND 	&	15.02.2017	&	57800.275	&	12.48	&	13.58	&	-	&	Sch 	&	FLI	\\
24.08.2010	&	55432.586	&	12.43	&	13.52	&	-	&	1.3-m	&	AND 	&	16.02.2017	&	57801.305	&	12.47	&	13.58	&	-	&	Sch 	&	FLI	\\
25.08.2010	&	55433.594	&	12.40	&	13.49	&	-	&	1.3-m	&	AND 	&	28.02.2017	&	57813.257	&	12.44	&	13.54	&	-	&	Sch 	&	FLI	\\
20.09.2010	&	55459.636	&	12.36	&	13.45	&	14.47	&	1.3-m	&	AND 	&	04.03.2017	&	57817.268	&	12.44	&	13.53	&	-	&	Sch 	&	FLI	\\
30.10.2010	&	55499.524	&	12.32	&	13.44	&	14.51	&	2-m	&	VA	&	01.04.2017	&	57845.249	&	12.45	&	13.57	&	-	&	Sch 	&	FLI	\\
31.10.2010	&	55500.503	&	12.36	&	13.51	&	14.61	&	2-m	&	VA	&	02.04.2017	&	57846.255	&	12.45	&	13.58	&	-	&	Sch 	&	FLI	\\
01.11.2010	&	55501.501	&	12.40	&	13.53	&	14.62	&	2-m	&	VA	&	17.09.2017	&	58013.548	&	12.47	&	13.59	&	-	&	Sch 	&	FLI	\\
01.11.2010	&	55501.546	&	12.36	&	13.46	&	-	&	Sch 	&	FLI	&	17.10.2017	&	58043.523	&	12.49	&	13.57	&	-	&	Sch 	&	FLI	\\
02.11.2010	&	55502.569	&	12.40	&	13.52	&	14.61	&	2-m	&	VA	&	18.10.2017	&	58044.595	&	12.43	&	13.51	&	-	&	Sch 	&	FLI	\\
04.11.2010	&	55505.472	&	12.32	&	13.41	&	-	&	Sch 	&	FLI	&	22.11.2017	&	58080.483	&	12.47	&	13.55	&	-	&	Sch 	&	FLI	\\
06.11.2010	&	55506.501	&	12.35	&	13.43	&	-	&	Sch 	&	FLI	&	23.11.2017	&	58081.486	&	12.48	&	13.58	&	-	&	Sch 	&	FLI	\\
07.11.2010	&	55507.511	&	12.32	&	13.41	&	-	&	Sch 	&	FLI	&	25.12.2017	&	58113.402	&	12.48	&	13.56	&	-	&	Sch 	&	FLI	\\
01.01.2011	&	55563.420	&	12.55	&	13.74	&	-	&	Sch 	&	FLI	&	26.12.2017	&	58114.397	&	12.46	&	13.54	&	-	&	Sch 	&	FLI	\\
06.01.2011	&	55568.388	&	12.44	&	13.57	&	14.68	&	2-m	&	VA	&	03.09.2018	&	58364.578	&	12.50	&	13.62	&	-	&	Sch 	&	FLI	\\
08.01.2011	&	55570.273	&	12.42	&	13.52	&	14.63	&	2-m	&	VA	&	05.11.2018	&	58428.435	&	12.50	&	13.61	&	-	&	Sch 	&	FLI	\\
09.01.2011	&	55571.378	&	-	&	13.53	&	14.65	&	2-m	&	VA	&	12.01.2019	&	58496.316	&	12.49	&	-	&	-	&	Sch 	&	FLI	\\
06.02.2011	&	55599.321	&	12.37	&	13.46	&	-	&	Sch 	&	FLI	&	28.02.2019	&	58543.232	&	12.56	&	-	&	-	&	Sch 	&	FLI	\\
07.02.2011	&	55600.299	&	12.34	&	13.40	&	-	&	Sch 	&	FLI	&	01.03.2019	&	58544.250	&	12.72	&	13.94	&	-	&	2-m	&	AND 	\\
04.04.2011	&	55656.253	&	12.36	&	13.46	&	-	&	Sch 	&	FLI	&	12.08.2019	&	58707.587	&	12.58	&	-	&	-	&	Sch 	&	FLI	\\
09.04.2011	&	55661.260	&	12.39	&	13.50	&	14.61	&	2-m	&	VA	&	01.09.2019	&	58727.587	&	12.77	&	-	&	-	&	2-m	&	AND 	\\
11.09.2011	&	55815.564	&	12.44	&	13.53	&	14.59	&	1.3-m	&	AND 	&	03.09.2019	&	58729.589	&	12.89	&	-	&	-	&	2-m	&	AND 	\\
12.09.2011	&	55816.591	&	12.43	&	-	&	14.57	&	1.3-m	&	AND 	&	02.10.2019	&	58758.539	&	12.53	&	-	&	-	&	Sch 	&	FLI	\\
20.09.2011	&	55824.537	&	12.40	&	13.50	&	14.60	&	1.3-m	&	AND 	&	15.01.2020	&	58864.358	&	12.49	&	-	&	-	&	Sch 	&	FLI	\\
08.10.2011	&	55842.510	&	12.40	&	13.49	&	14.54	&	1.3-m	&	AND 	&	16.01.2020	&	58865.361	&	12.49	&	-	&	-	&	Sch 	&	FLI	\\
14.10.2011	&	55848.501	&	12.42	&	13.51	&	14.53	&	1.3-m	&	AND 	&	18.01.2020	&	58867.372	&	12.60	&	13.80	&	-	&	2-m	&	AND 	\\
01.11.2011	&	55866.521	&	-	&	-	&	14.64	&	2-m	&	VA	&	20.01.2020	&	58869.361	&	12.61	&	-	&	-	&	2-m	&	AND 	\\
26.11.2011	&	55892.474	&	12.47	&	13.56	&	14.63	&	2-m	&	VA	&	21.01.2020	&	58870.366	&	12.42	&	13.61	&	-	&	Sch 	&	FLI	\\
27.11.2011	&	55893.417	&	12.40	&	13.49	&	-	&	Sch 	&	FLI	&	29.02.2020	&	58909.278	&	12.45	&	-	&	-	&	Sch 	&	FLI	\\
29.11.2011	&	55895.486	&	12.35	&	13.43	&	-	&	Sch 	&	FLI	&	21.03.2020	&	58930.318	&	12.47	&	-	&	-	&	Sch 	&	FLI	\\
30.11.2011	&	55896.443	&	12.38	&	13.50	&	-	&	Sch 	&	FLI	&	13.09.2020	&	59105.602	&	12.44	&	-	&	-	&	Sch 	&	FLI	\\
29.12.2011	&	55925.450	&	12.35	&	13.43	&	-	&	Sch 	&	FLI	&	22.11.2020	&	59176.451	&	12.40	&	13.62	&	-	&	Sch 	&	FLI	\\
01.01.2012	&	55928.381	&	12.39	&	13.48	&	-	&	Sch 	&	FLI	&	05.01.2021	&	59220.339	&	12.58	&	13.71	&	14.78	&	2-m	&	AND 	\\
16.03.2012	&	56003.258	&	12.34	&	13.42	&	-	&	Sch 	&	FLI	&	04.02.2021	&	59250.350	&	12.54	&	13.70	&	14.76	&	2-m	&	AND 	\\
26.03.2012	&	56013.259	&	12.41	&	13.52	&	14.58	&	2-m	&	VA	&	11.09.2021	&	59468.598	&	12.41	&	13.61	&	14.71	&	Sch 	&	FLI	\\
28.03.2012	&	56015.310	&	12.47	&	13.60	&	-	&	2-m	&	VA	&	30.11.2021	&	59549.451	&	12.41	&	-	&	-	&	Sch 	&	FLI	\\
21.08.2012	&	56160.582	&	12.43	&	13.49	&	-	&	1.3-m	&	AND 	&		&		&		&		&		&		&		\\
		\hline \hline
\end{longtable}}

\begin{figure}[]
	\begin{center}
		\centering{\epsfig{file=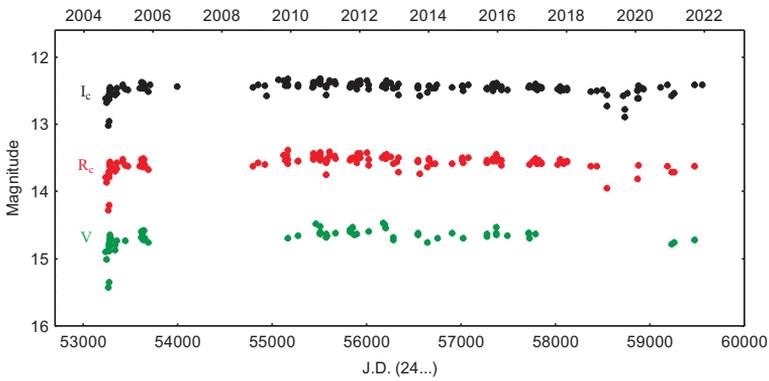, width=10cm}}
		\caption[]{$V(RI)_{c}$ light curves of LkH$\alpha$ 301 for the period August 2004$-$November 2021.}
		\label{fig4}
	\end{center}
\end{figure}

The measured color indices $V-I_{c}$ and $V-R_{c}$ versus the $V$ magnitude of LkH$\alpha$ 301 are plotted in Fig. 5.
A clear dependence can be seen from the figure: the star becomes redder as it fades.
This is an expected color variation in the TTSs.

\begin{figure}
	\begin{center}
		\includegraphics[width=6cm]{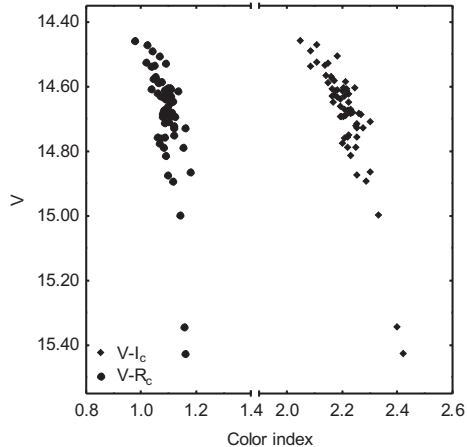}
		\caption{Color indices $V-I_{c}$ and $V-R_{c}$ versus the $V$ magnitude of LkH$\alpha$ 301.}\label{Fig5}
	\end{center}
\end{figure}

We utilized the software package \textsc{period04} (Lenz \& Breger 2005) to search for periodicity in the light variations of V2764 Ori and LkH$\alpha$ 301.
We did not identify any periodicity in their photometric behavior.
The reason for this negative result is probably the short life of the hot spots on the stellar surface.
Another reason may be the insufficient number of photometric points we have for the stars.
Due to negative declination, the field of the McNeil's Nebula can be observed from southern Europe only in the period August-April.
In some months and years, we have only a few observations of this field.

\section*{4. Concluding remarks}

We presented and discussed the optical CCD light curves and color$-$magnitude diagrams of V2764 Ori and LkH$\alpha$ 301.
Our observations cover 17 years and represent the first long-term $V(RI)_{c}$ monitoring of the stars.
The shape of the light curves, the brightness variations, and the observed amplitudes of both stars are typical for CTTSs.
Our study adds two CTTSs with long-term photometry to the family of known PMS stars.
We plan to continue our monitoring of the field of the McNeil's Nebula during the next years.

\section*{Acknowledgements}

This research has made use of NASA's Astrophysics Data System Abstract Service.
The authors thank the Director of Skinakas Observatory Prof. I. Papamastorakis and Prof. I. Papadakis for the award of telescope time.
This work was partly supported by the National Science Fund of the Ministry of Education and Science of Bulgaria under grand DN 18-13/2017 and by the research fund of the University of Shumen.

\end{document}